\documentclass{jpsj3}
\usepackage{txfonts}

\title{\textbf{A density-matrix renormalization group Study of one-dimensional
incommensurate quantum Frenkel-Kontorova model}}

 \author{Yongjun Ma$^1$, Jiaxiang Wang$^1$\footnote{Corresponding author: jxwang@phy.ecnu.edu.cn}, Xinye Xu$^1$, Qi Wei$^1$, Sabre Kais$^2$,$^3$}

\inst{$^1$State Key Laboratory of Precision Spectroscopy and
Department of Physics, East China Normal University, Shanghai
200062, China \\
$^2$Departments of Chemistry and Physics, Purdue University,
West Lafayette, Indiana 47907, USA \\
$^3$ Qatar Environment
and Energy Research Institute, Qatar Foundation, Doha, Qatar} 

\abst{
In this paper, the one-dimensional incommensurate quantum Frenkel-Kontorova model is investigate by a density-matrix renormalization group
algorithm. Special attention is given to the entanglement and the ground state energy. The energy gap between the ground state and the first
excited is also calculated. From all the numerical results, we have observed an obvious property changes from the pinned state to the sliding one as the quantum fluctuation is increased. But no expected quantum critical point can be justified by the present data. }

\begin{document}
\maketitle

\section{Introduction}
The Frenkel-Kontorova (FK) model describes a chain of
interacting particles in the presence of an external periodic
potential, which first appeared in 1938 \cite{FK1,FK2,FK3}. As a
discrete model, the classical FK model is non-integrable and has
been used as a generic tool to study many nonlinear effects such as
chaos, kinks and breathers since 1970's \cite{FKmodel1}.
\par
  Besides nonlinearity, another important feature of this model is the competition between two length scales: one
is the average distance between the neighboring particles and the
other is the length of the spacial period of the external potential. This competition can lead to quite interesting phenomena.
For example, if the ratio of the two length scales is a rational number, the system is said to
be commensurate and is always in a pinned state. Otherwise, it is incommensurate and there is a threshold $K_c$. For $K>K_c$,  the particles are
pinned and for $K<K_c$, they are depinned and can slide along the external potential \cite{FKmodel1}.
\par

The above scenario will be changed once we go to the quantum regime. Intuitively, it should be expected that, for a classically pinned state,
if the quantum fluctuation is high enough, it will get depinned and become a sliding one. This has been numerically testified as early as 1989 by Borgonovi \cite{Borgonovi} for
the incommensurate case, where the classical ground state is characterized by the fractal devil¡¯s staircase. Since then, there have been some papers devoted to this problem by Monte Carlo or variation methods \cite{F-K3,Borgonovi1,Berman1,Berman2,Hu1,Hu2,Ho,Hu3}. But due to the complexities related to the many body calculations, the corresponding investigations progress slowly. Since 2001, there are only a few papers devoted to this problem. For example, in 2003, Zhirov explored the tunneling properties by using path integral Monte Carlo method and put forward a "instanton glass state", which can change to a sliding state through a second-order quantum phase transition (QPT)\cite{F-K6}. In 2006, we developed a density-matrix renormalization group (DMRG) algorithm upon quantum FK model and have obtained much cleaner numerical results about the saw-tooth map, the coordinate correlations and the delocalization effect \cite{F-K4}. Through all these work, it has been shown that there are at least two kinds of different states for the quantum FK model. One is the pinned state and the other is the sliding state. The remaining question is what is the other properties of these states and how they help to clarify the essence of the the phase transitions.
\par
In this paper, we will use the quantum entanglement, the ground state energy and the energy gap between the ground state and the first excited state to explore more deeply how the state changes as we increase the quantum fluctuations. As we know, the entanglement is often taken as a physical resource for quantum communications and computations. Another well-known characteristic of entanglement lies in its critical behavior near QPT point \cite{entangle0}. Since 2000, there have been a lot of work demonstrating the application of entanglement as an auxilary signature of QPT \cite{entangle2,entangle3,entangle4,entangle5,entangle6}. Moreover, the energy of the ground state and the low excited states are also very important in understanding the QPT. Generally, these quantities are not easy to be calculated for the quantum FK model because the number of classcially excited equilibrium configurations is very huge and the band gap
is exponentially small as the system size is increased \cite{F-K7,F-K8,F-K5,F-K6}. But as we know, through the development in the past two decades since 1992 \cite{dmrg01}, DMRG has become one of the most efficient methods in dealing with the low-dimension strongly correlated quantum systems \cite{dmrg2}. Hence, we will use the DMRG algorithm to do the calculations as in Ref.\cite{F-K4}.
\par
In the following, we will first briefly review the model in order to introduce the notations and gives the main ideas of DRMG method to make the paper self-contained. Then the numerical results of entanglement and energies will be presented and analyzed. The final section is the summary.

\section{Model}

\begin{figure}\centering
\includegraphics[angle=-90,width=10cm]{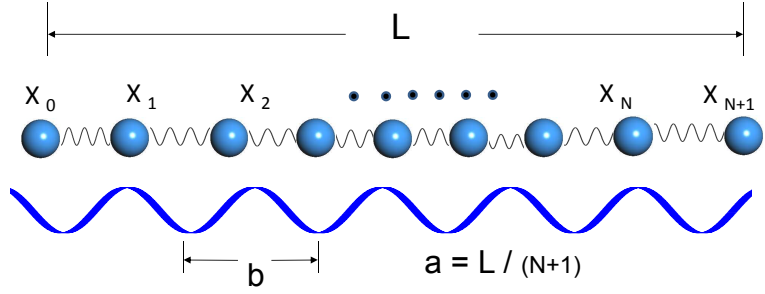}
\hfill \caption{Schematic diagram of the FK
model with finite size. A chain of $N+2$ particles connect by springs is subjected to
the action of an external periodic potential. The length of the chain
is $L$. The average distance between neighboring particles and the external potential period are
$a=\frac{L}{N+1}$  and $b$ respectively. $a/b= (\sqrt{5}-1)/2\approx 1/2,2/3,3/5,\cdots$. } \label{model}
\end{figure}

  Fig. (\ref{model}) presents a schematic diagram for the one-dimensional
quantum FK model. The total number of particles is $N+2$. By using the fixed boundary conditions with $x_{0}=0$, $x_{N+1}=(N+1)a$, we can express the quantum Hamiltonian as,
\begin{equation}
\hat{H}=\sum_{i=1}^{N}[- \frac{\hbar^2}{2m}\frac{\partial^2
}{\partial x_{i}^{2}}+ \frac{k}{2}( \hat x_{i+1} - \hat x_{i} -a)^2
-V\cos(\frac{2\pi}{b} \hat x_{i} )],     \label{E_interna0}
\end{equation}
where $\hbar$ is the Planck constant, $k$ is the elastic
constant and $V$ is the potential amplitude. $a$ and $b$ are the average distance between the particles and the spacial period of the external potential, respectively.
  \par
 By defining the wave number of
the external potential, $q_{0}=2\pi/b$, we can introduce the following dimensionless
parameters,
\begin{eqnarray}
  {{\hat X}_i} = {q_0}{{\hat x}_i},\mu  = {q_0}a,{{\hat U}_i} = {{\hat X}_i} - i\mu , \hfill \nonumber\\
  K = V\frac{{q_0^2}}{k},\tilde \hbar  = \frac{{q_0^2}}{{\sqrt {mk} }}\hbar ,\hat H' = \frac{{q_0^2}}{k}\hat H. \hfill \nonumber\\
\end{eqnarray}
 Then Eq. (\ref{E_interna0}) can be rewritten in a dimensionless form,
\begin{equation}
\hat{H'}=\sum_{i=1}^{N}[- \frac{\tilde{ \hbar}^2}{2}\frac{\partial^2
}{\partial U_{i}^{2}}+ \frac{1}{2}({\hat U_{i+1}}-{\hat U_{i}})^2
-K\cos(\hat{U_{i}}+i\mu)]. \label{reaction2}
\end{equation}

\par
  Next, to rewrite the above Hamiltonian in a
 second-quantized form, the following transformations are used,
 \begin{eqnarray}
  \hat{U_{i}} = \dfrac{1}{\sqrt{2}} \dfrac{\sqrt{ \tilde{\hbar} }}{\sqrt[4]{2}}({\hat a_{i}}^{\dag}+{\hat a_{i}}), \hat{P_{i}}= \dfrac{\partial }{\partial{U_{i}}} =
-\dfrac{1}{\sqrt{2}} \dfrac {\sqrt[4]{2}}{\sqrt{\tilde{\hbar}}}
({\hat a_{i}}^{\dag}-{\hat a_{i}}).
\end{eqnarray}
where $\hat{a_{i}}^{\dag}$ and $\hat{a_{i}}$ are the annihilation
 and creation operators satisfying $[\hat a_{i} ,\hat a_{j}^{\dag} ]=
 \delta_{ij}$,  $[\hat a_{i} ,\hat a_{j} ]= 0$,
 $[\hat a_{i} ^{\dag},\hat a_{j}^{\dag} ]= 0$.
Then,
\begin{eqnarray}
\hat{H'}& = & \sqrt{2} \tilde{\hbar}  \sum_{i=1}^{N} \{({\hat
a_{i}}^{\dag}{\hat a_{i}}
 +\frac{1}{2} )  -\frac{K}{\sqrt{2} \tilde{\hbar} }\cos[\frac{1}{\sqrt{2}} \frac{\sqrt{\tilde{ \hbar}} }
{\sqrt[4]{2}} (\hat a_{i} ^{\dag}+\hat a_{i} )+i\mu]\}\nonumber\\
  &  &
  -\sqrt{2}\tilde{ \hbar} \sum_{i=1}^{N}(\hat a_{i} ^{\dag} +\hat a_{i})(\hat a_{i+1}^{\dag} +\hat a_{i+1}).
    \label{action4}
\end{eqnarray}
 \par
It is obvious that there are three independent parameters in the above equations, namely $\hbar$ ,$K$ and $\mu$. The first one is the effective Plank
constant, denoting the quantum fluctuation. The second measures the strength of external potential. These two scaleless effective parameters can be
varied by changing the elastic constant, particle mass and the magnitude of external potential. The third independent parameter $\dfrac{\mu}{2\pi}= a/b $ denotes the commensurability property of system. In this paper, we are interested in the incommensurate case and take $\dfrac{\mu}{2\pi}$ to be the famous golden mean ratio, $(\sqrt{5}-1)/2$. Numerically, it is impossible to realize a true irrational number on the digital computers. So we will use Fibonacci series to approximate $(\sqrt{5}-1)/2$, which means,
\begin{equation}
\frac{a}{b}=\frac{\frac{L}{b}}{N+1}=\frac{\sqrt{5}-1}{2}\approx \frac{N_1}{N_2}=\frac{2}{3},\frac{3}{5},\frac{5}{8},\frac{8}{13},\frac{13}{21},\frac{21}{34},\frac{34}{55},\cdots,
\end{equation}
where $L=N_1b=N_2a$ is the length of the chain.
Hence for a finite size system, we will take $N=N_2-1$ systematically in order to obtain convergent results, i.e. $N=1,2,4,7,12,20,33,54,\cdots$. In this way, it can be guaranteed that if all the particles are evenly distributed along the external potential, none will occupy the same location after the chain is folded into the region of one spacial period except the two particles on the boundary. Now we are in a good starting position to carry out the DMRG algorithm with fixed boundary condition, which makes it extremely convenient to carry out DMRG algorithm with fixed boundary condition. As to the concrete steps of the algorithm realization on quantum FK model, they has been well explained in Ref. \cite{F-K4}. In the next section, we will present the main steps to realize the DMRG algorithm in order to make the paper self-contained.
\section{DMRG Method}
\begin{figure}\centering
\includegraphics[angle=-90,width=12cm]{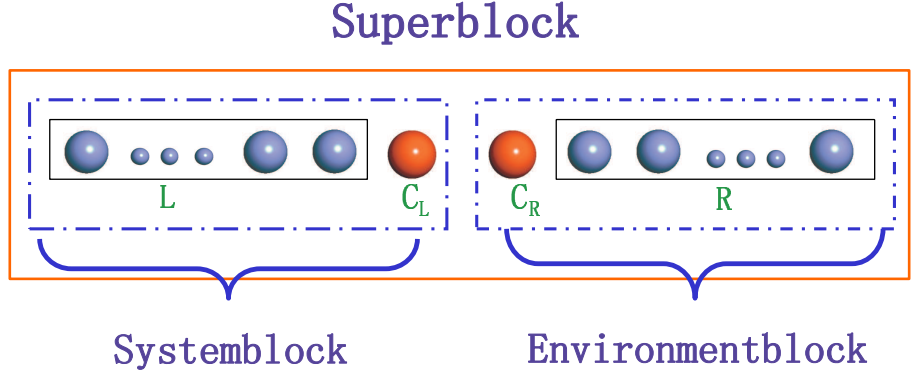}
\caption{Schematic explanation of DMRG algorithm.}
\label{DMRG}
\end{figure}
The main idea of the DMRG algorithm is to truncate the Hilbert space by the reduced density matrix calculated from the targeted wave function\cite{dmrg1}. We will use Fig. \ref{DMRG} to explain how this technique works. Firstly, let us cut the Hilbert space dimension for each particle to be $n$ with the bare bases $\phi_i(i=1,\cdots,n)$ to be the $n$ energy levels of $H_{0i}$, i.e.
\begin{eqnarray}
{\hat H'_{i0}}  =  \sqrt 2 \tilde{ \hbar} \left( {{{\hat a}_i}^\dag {{\hat a}_i} + \frac{1}{2}} \right)- {K} \cos\left[ {\frac{1}{{\sqrt 2 }}  \frac{\sqrt{\tilde{ \hbar}} }   {{\sqrt[4]{2}}}\left({\hat a_i^\dag  + {{\hat a}_i}}\right)+ i\mu }\right].
\end{eqnarray}
As a starting point, the effective bases for the left and right blocks shown in Fig. \ref{DMRG} are assumed to have been found as $\psi_i^L$ and $\psi_i^R (i=1,\cdots,n)$, respectively. Our aim is to find $n$ optimal bases for the system block composed of the left block $L$ and the left central particle $C_L$ with the environment block composed of the right block $R$ and the right central particle $C_R$. Once this aim is realized, we can repeat this process to expand the new left blocks by one site each time until it includes all the left $N-2$ particles. Then by swopping the role of the system block and the environment one, the same process will be repeated from the right-hand side to the right. In this way, we can obtain the updated optimal bases for all the right blocks under the new environment blocks. This sweeping process will be iterated until the required convergence is reached.

The key to the success of DMRG algorithm is how to find the $n$ optimal bases for the system block. Take the ground-state wave function as an example. It has been proved \cite{dmrg01} that the $n$ optimal bases can be chosen as the eigenfunctions of the following reduced density matrix with the largest $n$ eigenvalues,
\begin{equation}
{\rho _{system{\rm{ }}block}} = Tr\left( {\left| \psi  \right\rangle \left\langle \psi  \right|} \right),
\end{equation}
where $Tr$ denotes the trace over bases of the environment blocks $\left| {\phi _i^R} \right\rangle \left| {\psi _j^R} \right\rangle (i,j = 1, \cdots ,n)$ and ${\left| \psi  \right\rangle }$ is the targeted ground state wave function of the superblock Hamiltonian composed of both the system and the environment blocks. After the $n$ optimal bases are acquired, the dimension of the system block will be truncated from $n^2$ to $n$. The error will be proportional to the sum of the omitted eigenvalues of the reduced density matrix ${\rho _{system{\rm{ }}block}}$. Normally it is quite small. In FK model, it can be kept below $10^{-10}$ if $n=8$ is used. Once the $n$ optimal bases are decided, they will be used to reconstruct all the operators of the system block and also the new Hamiltonian in the next step. The energy gap is calculated similarly so long as we target both the ground state and the first excited wave functions at the same time.

The above procedures are generic to all the DMRG algorithm. When applying them to FK model, we need to make two revisions in order to deal with the huge Hilbert space related to the bosonic system. One is the omission of the central particle $C_R$ because the calculation is too intensive if it is included. The other is related to the truncation of the local Hilbert space of $C_L$. A feeding process based upon higher energy levels has been designed to improve the truncation efficiency. It has been well explained in Ref. [\cite{F-K4}] and will not be repeated here.

\section{RESULTS AND DISCUSSION}
\subsection {Entanglement}
\begin{figure}\centering
\includegraphics[angle=-90,width=14cm]{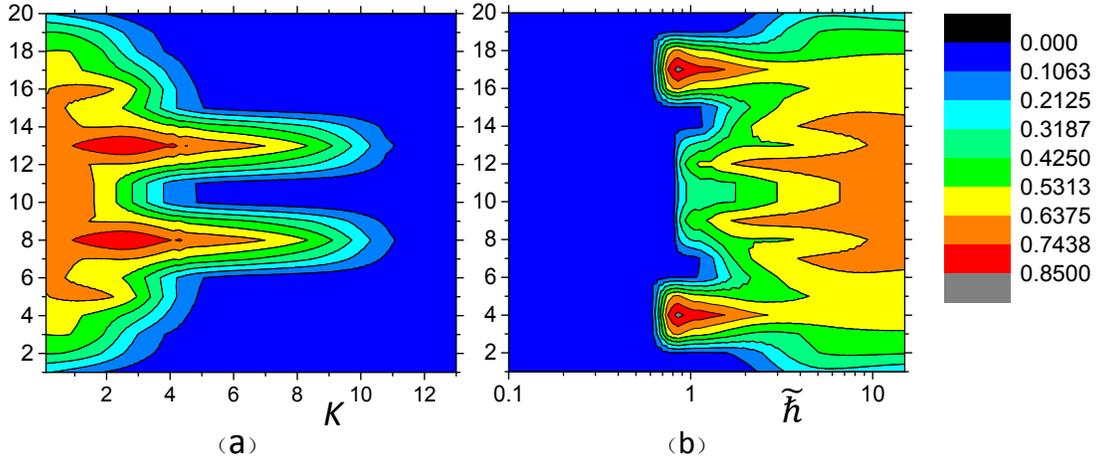}
\caption{Variations of the single-site entanglement against the external potential $K$ with $\hbar=1$ in (a) and the quantum fluctuation $\hbar$ with {K=5} in (b). The ordinate axis denotes the ordinal number of the particles. For the system size, $N=20$.} \label{entangle1}
\end{figure}
\par
  Quantum entanglement, characterized  as a quantum superposition,
is a miracle in modern physics. It is considered as a potential
resource which has been widely applied in quantum communications and
quantum information processings. In QPT, it can be taken as an order parameter.
\par
 To provide a quantitative measure of the entanglement here, we will use the von
Neumann entropy, which is defines as,
\begin{equation}
S  =  -tr(\rho\ln \rho ),
\end{equation}
where $\rho$ is the density matrix for the interested state
 and $tr$ denotes the trace. In our work, the single-particle entanglement will be calculated, which is obtained by using,
\begin{equation}
S_i  =  -tr(\rho_i\ln \rho_i ),
\end{equation}
in which $\rho_i=Tr_i\left | \Psi\right\rangle \left \langle \Psi\right |$ with $\left | \Psi\right\rangle$ to be the
ground state wave function calculated by DMRG method. The subscription in $Tr_i$ means the tracing over all the particle space except the $i$th one.

We first fix the parameter $\tilde{\hbar}=1$ and then study the entanglement dependence upon $K$.
The results are given in  Fig.\ref{entangle1} (a). It is obvious that the
entanglement gets smaller and smaller with the increase of $K$. This vividly demonstrates the pinning property of the external potential. As we know, for $\tilde{\hbar}=0$, FK model reduces
to a chain of harmonic oscillators and each particle is entangled with the rest of the system. As a whole, the harmonic chain can slide freely. Once the external potential is imposed, the particles will tend to be locked in the potential valley and at the same time, the entanglement will decrease. When the entanglement approaches zero, the whole system will change to a pinned state. In classical regime with $\hbar=0$, there is a clear transition point with $K_c=0.971635\cdots$, called Transition by Breaking of Analyticity (TBA) \cite{chaos1,chaos2,chaos3}. Here in quantum regime, with $\tilde{\hbar}=1$, this transition point seems to be smeared out by the quantum fluctuations. We can only observe a continuous state change around $K=5\sim 7$. Compared with the classical threshold values $K_c=0.971635\cdots$ to lock the particles, this means higher external potential is needed to pin the chain.

The results for the influence of $\tilde{\hbar}$ upon the entanglement with $K=5$ are shown in Fig.\ref{entangle1} (b). Classically, the system is in a pinned state and no entanglement exists. As we increase $\tilde{\hbar}$, the quantum fluctuations will help to entangle different parts of the chain. Once $\tilde{\hbar}$ approaches some threshold value, all the particles will be entangled to the rest of the system.  This is quite similar to the particle correlations studied in Ref. \cite{F-K4}. It is quite understandable since entanglement is just a special kind of quantum correlation. To check the critical properties of the transition from pinned state to sliding state, we have tried to make use of the average entanglement,
\begin{equation}
{S_E} = \frac{1}{N}\sum\limits_{i = 1}^N {{S_i}}.
\end{equation}
 It is found that $S_E$ varies continuously with $\hbar$. But unfortunately, with finite-size scaling technique as we have used in Ref.\cite{entangle3}, no collapse with different sizes can be acquired. That means no convergence of $S_e$ with $N$ can be acquired. The transition is more like a crossover around $\tilde{\hbar}\approx 1\sim 2$, i.e. the system seems to go through a continuous "melting" process from pinned state to a sliding one with no critical point. This is in contrast to the Zhirov's finding \cite{F-K6}. He claims the transition to be a continuous critical phase transition. Since we are using different order parameters, we still need more work to verify the essence of this transition.

Anyway, the quantum fluctuation and the external potential are two competing factors influencing the local entanglement. They will decide together whether the system is in a pinned or sliding state.

\subsection {Ground state energy}
\begin{figure}\centering
\includegraphics[angle=-90,width=12cm]{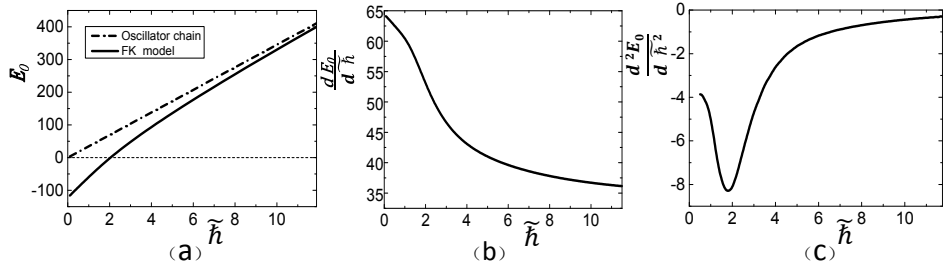}
\caption{(a) Dependence of the ground state energy upon the quantum fluctuation with $K=5$. For contrast, the results for a harmonic chain is also given as a solid line.
 (b) and (c) give the first-order and second-order derivative of ground state energy with respect to $\hbar$, respectively. For system size, $N=54$. } \label{Eground1}
\end{figure}
For a classical system, all physical quantities can be precisely determined simultaneously. But
for a quantum mechanical system, they have to observe Heisenberg's uncertainty principle, which will make the
quantum state have some novel properties. For example, the energy of a quantum harmonic
oscillator in the ground state is not zero at all due to the quantum fluctuations. The same thing happens in quantum FK model.
Classically, in a pinned state, the particles are localized by the
Peierls-Nabarro potential and the ground energy is always
negative if the correspond harmonic chain energy is set as zero. But if the
quantum fluctuations is introduced, the ground state energy will increase with $\hbar$. During this process, the particles could overcome the Peierls-Nabarro barrier at some threshold value and become depinned. Hence the ground state energy is also a very important quantity to demonstrate the state transition from pinned state to sliding one.

\begin{figure}
\centering
\includegraphics[angle=-90,width=14cm]{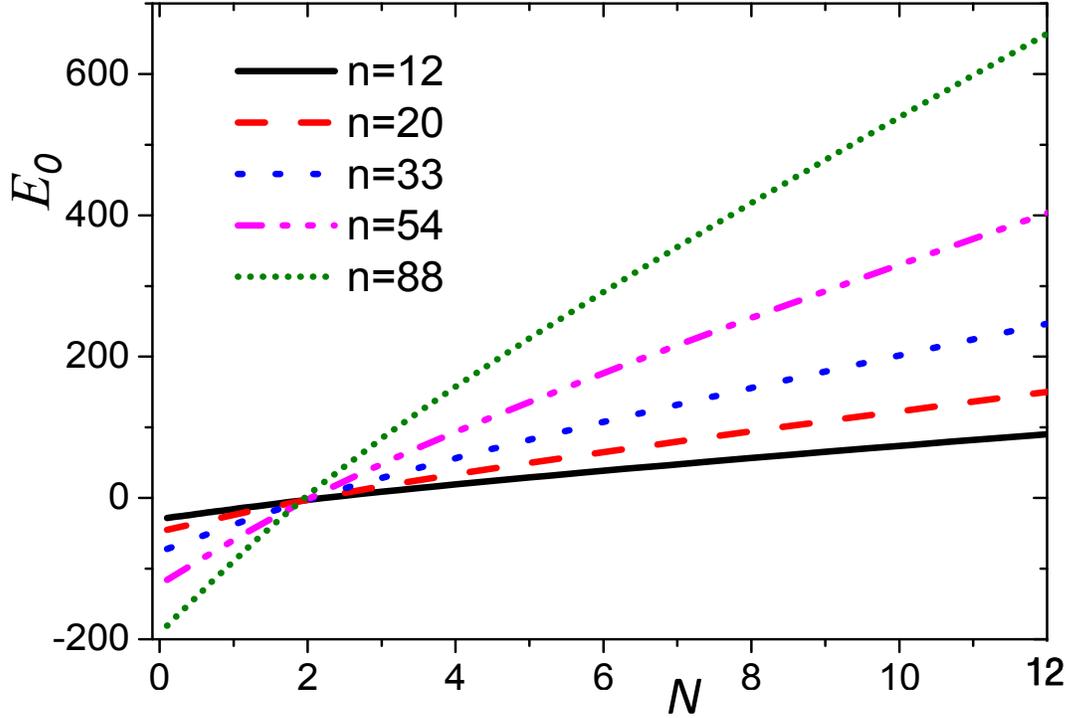}
\caption{ (a)The same as in Fig. (\ref{Eground1}) but for different system sizes.(b)Dependence of the energy gap between the ground state and the first excited state upon the quantum fluctuation for different system sizes. The legend denotes the total number of sits or particles. The parameter is   $K=5$.} \label{Eground2}
\end{figure}

    Fig.\ref{Eground1}(a) gives the dependence of the ground state
energy $E_{0}$ of both the FK model and an harmonic chain upon $\tilde{\hbar}$. From the figure, we can
see two quantum effects. Firstly, $E_{0}$ increases with $\tilde{\hbar}$. This is easily understandable due to the increase of the
quantum-fluctuation-induced kinetic energy. Secondly, the
energy difference gets
smaller and smaller with the accretion of the quantum fluctuation, which implicates that the FK
model behaves more and more like an effective harmonic chain.  It is also interesting to note that the ground state energy of the FK model will be
zero at a point near $\tilde{\hbar}\approx 2$, which is also the transition point obtained by Zhirov \cite{F-K6} for the transition from instant glass state to a sliding one. To get more information around this point, we also
calculates the first and second derivative of $E_{0}$ with respect
to $\tilde{\hbar}$ in Fig. \ref{Eground1}(b) and (c), respectively. It is obvious that in Fig. \ref{Eground1}(c) there is a minima at $\tilde{\hbar}\approx 2$, which can be regarded as a signature of a phase change from pinned state to a sliding one.
\par
 In order to understand the physical properties of the above-mentioned phase change, we have also made the calculation of the ground state energy with different system sizes. The results are displayed in Fig.\ref{Eground2}(a). It is marvelous to see that all lines pass though the same point at $\tilde{\hbar}\approx 2$. This gives strong implications over the types of the phase changes to be a continuous QPT. But unfortunately once again, no matter how hard we try with the finite-size scaling technique, these curves just don't collapse onto one curve. So we are still quite hesitated to claim the phase transition here as a critical phenomenon. To explore more deeply into this issue, next we will go to study the energy gap of the system.
\subsection {Energy gap}
\begin{figure}\centering
\includegraphics[angle=-90,width=12cm]{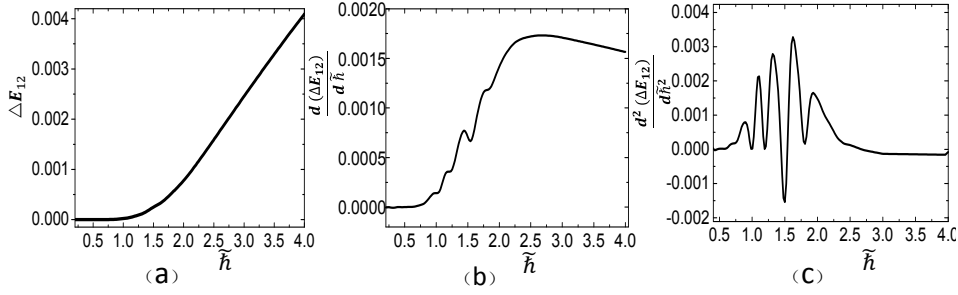}
\caption{(a)  and (b) give the first-order and second-order derivative of the energy gap with respect to $\tilde{\hbar}$ with $K=5$, respectively. For system size, $N=54$.  The inset in (a) depicts the dependence of
of the energy gap between the ground state and the first excited state upon the quantum fluctuation.   } \label{Egap}
\end{figure}
\par
 The structure of the energy levels in the incommensurate classical FK model is quite complicated since there are a
huge amount of meta stable configurations very close to the ground
state, especially for a long chain. As the band gap between the
ground state and the lowest excited state is extremely small, the FK
model will be unstable when it is perturbed. Hence the quantum
fluctuations can exert a strong influence on the classical energy levels. Luckily, DMRG can give a very precise calculation of the energy gap between the
ground state and the first excited state, although the amount of calculation is very huge.

\par
Fig.\ref{Egap} gives the energy gap $\Delta E_{12}$ together with its first- and second-order derivatives over $\hbar$.
As can be seen from Fig.\ref{Egap}(a),
the curve changes to a straight line when $\tilde{\hbar}>2$. The variation is just like that of a harmonic chain, whose
energy gap can be analytically expressed as $\bigtriangleup E_{12}= \tilde{\hbar}\cdot \sin
[\pi/(N+1)])$, which obviously has a linear dependence upon $\tilde{\hbar}$. This once again shows the similarity between the FK model in the sliding phase and the harmonic chain. Here again, no critical property of $\Delta E_{12}$ can be justified either by the finite-size scaling. But another phenomenon takes our attention, namely, in Fig.\ref{Egap}(c), the second-order derivative demonstrates strong oscillations between $\tilde{\hbar}\sim 1$ and $\tilde{\hbar}\sim 2$. Hence $\tilde{\hbar}\sim 1$ seems also to be a special point here. By looking backward to all the above numerical results together with the data we once got before \cite{F-K4}, we might be able to say that there potentially exist three phases: the pinned phase when $\tilde{\hbar}<1$, the sliding phase when $\tilde{\hbar } >2$ and a crossover phase when $1<\tilde{\hbar }<2$. And for the physical quantities we have investigated in this paper, no critical property can be justified.
\section{SUMMARY}
    In summary, we have investigated three physical
quantities of the quantum one-dimensional incommensurate
Frenkel-Kontorova model by a DMRG algorithm. They are the entanglement and energy of the ground state and the energy gap between the ground state and the first excited state. Their dependence
upon the quantum fluctuations are explored in details, which points to the existence of three kinds of phases: the conventional pinned and sliding phases and the crossover phase in between. For $K=5$, the transition happens at $\tilde{\hbar }\approx 1$ and $\tilde{\hbar} \approx 2$. By finite-size scaling technique, no critical property of the studied quantities can be justified in the present work. Our future work will focus upon looking for a more proper order parameter to describe the three phases. Morever by checking the dynamics of this model \cite{wang2007} might also help to elucidate the essence of these phase changes.

\section{ ACKNOWLEDGMENTS }
This work is supported by the National Natural Science Foundation of
China under Grant Nos. 11274117 and 11134003 and Shanghai Excellent academic leaders Program of China (Grant No. 12XD1402400).


\begin{thebibliography}{40}

\bibitem{FK1}    Ya. Frenkel, T.Kontorova,  Phys. Z. Sowietunion {\bf 13}, 1 (1938).
\bibitem{FK2}    T. A. Kontorova,  Ya. I. Frenkel, Zh. Eksp. Teor. Fiz   {\bf 8}, 89 (1938).
\bibitem{FK3}    T. A. Kontorova,  Ya. I. Frenkel, Zh. Eksp. Teor. Fiz   {\bf 8}, 1340 (1938).


\bibitem{FKmodel1}  M. Braun and Yuri S. Kivshar, The Frenkel-Kontorova Model:
   Concepts, Methods, and Applications (Springer, Berlin, 2003).
\bibitem{Borgonovi} F. Borgonovi, I. Guarneri, and D. L. Shepelyansky{\bf 63}, 2010 (1989).
\bibitem{F-K3}    B. Hu,  B. Li, W. Zhang,  Phys. Rev. E {\bf 58}, 4068 (1998).

\bibitem{Borgonovi1} F. Borgonovi, I. Guarneri and D. Shepelyansky, Z. Phys. B {\bf 79},133 (1990).
\bibitem{Berman1} G. P. Berman, E. N. Bulgakov and D. K. Campbell, Phys. Rev. B {\bf 49}, 8212 (1994).
\bibitem{Berman2} G. P. Berman, E. N. Bulgakov, and D. K. Campbell. Physica D {\bf 107}, 161 (1997)
\bibitem{Hu1} B. Hu and B. Li, Europhys. Lett. {\bf 46}, 655 (1999).
\bibitem{Hu2} B. Hu and B. Li, Physica A {\bf 288}, 81 (2000).


    \bibitem{Ho} C. L. Ho and C. I. Chou,  Phys. Rev. E {\bf 63}, 016203 (2000).
\bibitem{Hu3} B. Hu, B. Li and H. Zhao, Europhys. Lett.{\bf 53}, 342 (2001).

\bibitem{F-K6}    O. V. Zhirov,  G. Cassati, D. L. Shepelyansky,  Phys. Rev. E {\bf 65}, 026220 (2002).
\bibitem{F-K4}    B. Hu,  J. Wang,  Phys. Rev. B {\bf 73}, 184305 (2006).
\bibitem{entangle0} A. Osterloh, Luigi Amico, G. Falci,  Rosario Fazio, Nature {\bf 416}, 608 (2002).


\bibitem{entangle2} X. Wang, Phys. Rev. A {\bf 64}, 012313 (2001).
\bibitem{entangle3} J. X. Wang, Sabre Kais, Phys. Rev. A {\bf 70}, 022301 (2004).
\bibitem{entangle4} G. Vidal, J. I. Latorre, E. Rico, and A. Kitaev, Phys. Rev. Lett.{\bf 90}, 227902 (2003).
\bibitem{entangle5} J. I. Latorre, E. Rico, G. Vidal, Quant. Inf. Comp.  {\bf 4}, 48 (2004)
\bibitem{entangle6} S. Gu, S. Deng, Y. Li, H. Lin, Phys. Rev. Lett. {\bf 93}, 086402 (2004).
\bibitem{F-K7}    R. Schilling, Phys. Rev. Lett. {\bf53},2258 (1984).
\bibitem{F-K8}    P. Reicher
\bibitem{F-K5}    O. V. Zhirov,  G. Cassati, D. L. Shepelyansky,  Phys. Rev. E {\bf 67}, 056209 (2003).t R.Schilling, Phys.Rev.B {\bf 32},5731 (1985).



\bibitem{dmrg01}   S. R. White,zPhys. Rev. Lett. {\bf 69}, 2863 (1992).
\bibitem{dmrg2}    S. R. White , A. E. Feiguin, Phys. Rev. Lett. {\bf 93}, 076401 (2004).
\bibitem{dmrg1}    S. R. White, Phys. Rev. B {\bf 48}, 10345 (1993).


\bibitem{chaos1}   J. M. Greeme,  J. Math. Phys. {\bf 20}, 1183 (1979).
\bibitem{chaos2}   R. J. Shenker, L. P. Kadanoff, J. stat. Phys. {\bf 27}, 631 (1982).
\bibitem{chaos3}   J. M. Greeme, R. S. Mackay, F. Vivaldi, M. J. Feigenbaum, Physica D {\bf 3},486 (1981).


\bibitem{wang2007}  J. X. Wang, Bambi Hu and XiaoqunWang, Prog. Theo. Phys., Suppl. {\bf 166}, 95 (2007).


\end{thebibliography}
\end{document}